\documentclass[showpacs,preprintnumbers,amsmath,amssymb]{revtex4}

\usepackage{graphicx}
\usepackage{dcolumn}
\usepackage{bm}
\usepackage{color}

\begin{document}


\title{Charged Particle Motion Around
Rotating Black Hole in Braneworld\\ Immersed in Magnetic Field}

\author{Ahmadjon Abdujabbarov}%
   \email{abahmadjon@yahoo.com }
\author{Bobomurat Ahmedov}
 \email{ahmedov@astrin.uzsci.net}

 \affiliation{%
Institute of Nuclear Physics, Ulughbek, Tashkent
              100214, Uzbekistan\\
    Ulugh Beg Astronomical Institute,    Astronomicheskaya 33,
    Tashkent 100052, Uzbekistan\\
    Inter University Centre for Astronomy \& Astrophysics,
    Post Bag 4, Pune 411007,
    India}
\date{\today}
\begin{abstract}

Analytical solutions of Maxwell equations in background spacetime
of black hole in braneworld immersed in external uniform magnetic
field have been found. Influence of both magnetic and brane
parameters on effective potential of the radial motion of charged
test particle around slowly rotating black hole in braneworld
immersed in uniform magnetic field has been investigated by using
Hamilton-Jacobi method. Exact analytical solution for dependence
of the radius of the innermost stable circular orbits (ISCO)
$r_{\rm ISCO}$ from brane parameter for motion of test particle
around nonrotating isolated black hole in braneworld has been
derived. It has been shown that radius $r_{\rm ISCO}$ is
monotonically growing with the increase of module of brane tidal
charge. Comparison of the predictions on $r_{\rm ISCO}$ of the
brane world model and of the observational results of ISCO from
relativistic accretion disks around black holes provided upper
limit for brane tidal charge $\lesssim 10^9 {\rm cm}^2$.

\end{abstract}

\pacs{04.50.-h, 04.40.Dg, 97.60.Gb}
\maketitle

\section{\label{sec:intro}Introduction}

The idea that our Universe might be a three-brane \cite{rs99},
embedded in a higher dimensional spacetime, has recently attracted
much attention. For astrophysical interests, static and
spherically symmetric exterior vacuum solutions of the brane world
models were initially proposed by Dadhich et al \cite{nkd00,m04}
which have the mathematical form of the Reissner–-Nordstr\"{o}m
solution, in which a tidal Weyl parameter $Q^*$ plays the role of
the electric charge {squared} of the general relativistic
solution. The so-called DMPR solution was obtained by imposing the
null energy condition on the three-brane for a bulk having nonzero
Weyl curvature.

Observational possibilities of testing the brane world black hole
models at an astrophysical scale have intensively discussed in the
literature during the last years, for example through the
gravitational lensing~\cite{k06,k05,bozza02,ee05,wr06,pk07}, the
motion of test particles~\cite{i05} and  the classical tests of
general relativity (perihelion precession, deflection of light and
the radar echo delay) in the Solar system~\cite{bhl08}. The role
of the tidal charge in orbital models of high-frequency
quasiperiodic oscillations observed in neutron star binary systems
has been also studied ~\cite{kst08}. In the paper~\cite{harko08}
the energy flux, the emission spectrum and accretion efficiency
from the accretion disks around several classes of static and
rotating brane-world black holes have been obtained. The complete
set of analytical solutions of the geodesic equation of massive
test particles in higher dimensional space–times which can be
applied to braneworld models is provided in the recent
paper~\cite{hkkl08}. {Recently the deflection angle of light rays
caused by a massive black hole in braneworld  in the weak lensing
approach has been derived, up to the second order in perturbation
theory \cite{ref1,ref2}. }

A braneworld corrections to the charged rotating black holes and
to the perturbations in the electromagnetic potential around black
holes are studied in~\cite{ag05,rc06}. Our preceding
paper~\cite{af08} was devoted to the stellar magnetic field
configurations of relativistic stars in dependence on brane
tension. Here we plan to study electromagnetic fields and particle
motion around rotating black hole in braneworld immersed in
uniform magnetic field. The study of the particle orbits could
provide an opportunity for constraining the allowed parameter
space of solutions, and to provide a deeper insight into the
physical nature and properties of the corresponding spacetime
metrics. Therefore, this may open up the possibility of testing
brane world models by using astronomical and astrophysical
observations around black holes, in particular observationally
measured ISCO radii around black holes in principle may give
definite constraints on the numerical value of the brane tidal
charge.

The paper is organized as follows. In section \ref{sec:emfield} we
look for exact solutions of vacuum Maxwell equations in spacetime
of the rotating black hole in braneworld immersed in uniform
magnetic field. In the next section \ref{motion}  motion of
charged particles around black hole in braneworld immersed in
uniform magnetic field has been studied in slow rotation
approximation. We obtain the effective potential for any particle
with a specific angular momentum, orbiting around the black hole,
as a function of the magnetic field, and of the tidal charge of
the black hole. Exact expression for dependence of radius of
innermost stable circular orbit from brane charge has been found
in section \ref{branemotion} for the test particle moving in the
equatorial plane of the black hole in braneworld when both
rotation and magnetic parameters are neglected for the simplicity
of calculations. {Then we present clear derivation of the capture
cross section of slowly moving test particles by black hole in
braneworld.  The exact expressions for critical angular momentum
of the test particle and corresponding radius of particle unstable
circular orbits around the black hole have been presented.} For
different tidal charges, the values of the radii of the marginally
stable orbits around black hole in braneworld, are also plotted.
The conclusion and discussion of the obtained results can be found
in section \ref{conclusion}.

We use in this paper a system of units in which $c = 1$, a
space-like signature $(-,+,+,+)$ and a spherical coordinate system
$(t,r,\theta ,\varphi)$. Greek indices are taken to run from 0 to 3,
Latin indices from 1 to 3 and we adopt the standard convention for
the summation over repeated indices. We will indicate vectors
with bold symbols ({\it e.g.} ${\boldsymbol B}$) .

\section{\label{sec:emfield}
Rotating Black Hole in Braneworld Immersed in Uniform Magnetic Field}

Spacetime metric of the rotating black hole in braneworld in
coordinates ${t,r,\theta,\varphi}$ takes form {(see e.g,
\cite{ag05})}
\begin{equation}\label{metric}
ds^2=-\frac{\Delta-a^2\sin^2\theta}{\Sigma}dt^2+\frac{(\Sigma+a^2\sin^2
\theta)^2  -\Delta a^2 }{\Sigma}\sin^2 \theta d\varphi^2
+\frac{\Sigma} {\Delta}dr^2 +\Sigma d\theta^2
-2\frac{\Sigma+a^2\sin^2 \theta-\Delta}{\Sigma}a\sin^2 \theta
d\varphi dt\ ,\ \quad
\end{equation}
where $\Sigma=r^2+a^2\cos^2\theta$, $\Delta=r^2+a^2-2Mr+Q^*$,
$Q^*$ is the bulk tidal charge, $M$ is the total mass and $a$ is
related to the angular momentum of the black hole.

{It is not difficult to show that the electromagnetic corrections
created by the external magnetic field being proportional to the
electromagnetic energy density are rather small in most black
holes. Indeed if $B$ is the external magnetic field around black
hole in braneworld of total mass $M$ at radius $r$, these
corrections are at most
\begin{eqnarray}
&& \frac{B^2 r^3}{8\pi M c^2}\simeq 7\cdot10^{-4}
\left(\frac{B}{10^{12}\ {\rm G}}\right)^2 \left(\frac{10^6
M_{\bigodot}}{M}\right) \left(\frac{r}{1.5\cdot10^{6}\ {\rm
km}}\right)^3 .
\end{eqnarray}
}

Here we exploit the existence in this spacetime of a timelike
Killing vector $\xi^\alpha_{(t)}$ and spacelike one
$\xi^\alpha_{(\varphi)}$ being responsible for stationarity and
axial symmetry of geometry, such that they  {satisfy the} Killing
equations
$
 \xi_{\alpha ;\beta}+\xi_{\beta;\alpha}=0\ ,
$
and  {consequently the} wave-like equations (in vacuum spacetime)
$
\Box{\xi^\alpha}=0\ ,
$
which gives a right to write the solution of vacuum Maxwell
equations $\Box A^\mu=0$ for the vector potential $A_\mu$ of the
electromagnetic field in the Lorentz gauge in the simple form
$
 A^\alpha=C_1 \xi^\alpha_{(t)}+C_2
\xi^\alpha_{(\varphi)}\ . $ \cite{w74}
The constant $C_2=B/2$, where gravitational source is immersed in
the uniform magnetic field $\textbf{B}$ being parallel to its axis
of rotation. The value of the remaining constant $C_1=aB$ can be
easily calculated from the asymptotic properties of spacetime
(\ref{metric}) at the infinity (see e.g. our preceding
paper~\cite{aak08} for the details of typical calculations).

Finally the components of the 4-vector potential $A_\alpha$ of the electromagnetic
field will take a form
\begin{eqnarray}
&&
A_0=\frac{aB}{2\Sigma}\left[(2-\sin^2\theta)(a^2\sin^2\theta-\Delta)-
\Sigma\sin^2\theta\right]\ ,\  A_1=A_2=0\ , \nonumber
\\
&& \label{potentials}A_3=\frac{B\sin^2 \theta}{2\Sigma}
\left[(\Delta-\Sigma-a^2)(2-\sin^2 \theta)a^2
+\Sigma(\Sigma+\sin^2 \theta )\right]\ .
\end{eqnarray}

The nonvanishing orthonormal components of the electromagnetic fields measured
by zero angular momentum observers (ZAMO) with the four-velocity
components
\begin{equation}
(u^{\alpha})_{_{\rm ZAMO}}\equiv
    {\frac{K}
    {\sqrt{\Delta\Sigma}}}
    \ \left(1,0,0, \frac{\Sigma a^2\sin^2 \theta}{\Delta-a^2\sin^2
    \theta}-1
    \right) \ ,
(u_{\alpha})_{_{\rm ZAMO}}\equiv
    {\frac
    {\sqrt{\Delta\Sigma}}{K} }\ \big(1,0,0,0 \big) \
\end{equation}
are given by expressions
\begin{eqnarray}
\label{e1} && E^{\hat r} =  \frac{aB}{\Sigma^2}\bigg\{2(M-r)
+M\sin^2\theta+\frac{\sin^4 \theta}{\Delta-
a^2\sin^2\theta}(\Sigma-\Delta
+a^2\sin^2\theta)\bigg[r\Sigma+a^2(2-\sin^2\theta)\nonumber\\
&& \qquad \quad \times \frac{r\Delta-a^2r+(M-r) \Sigma}
{\Sigma}\bigg]+\frac{r}{\Sigma}(2-\sin^2\theta)
\big[\Sigma^2+(\Delta-a^2 \sin^2\theta)(2-\sin^2\theta)\big]
\bigg\}K,\\
\label{e2} && E^{\hat \theta}= \frac{aB\sin
2\theta}{2\Sigma^2\sqrt{\Delta}}
\bigg\{a^2\sin^2\theta-\Delta-\Sigma
+\frac{a^2\sin^2\theta-\Delta+\Sigma} {\Sigma}a^2(2-\sin^2 \theta)
+ \frac{a^2\sin^2\theta-\Delta+\Sigma}{\Delta\csc^2\theta- a^2}
\big[(\Sigma+a^2\nonumber\\
&&\qquad \quad \times(2+\cos^2\theta)-\Delta)a^2\sin^2
\theta+\Sigma(\Sigma+a^2\sin^2 \theta)-\frac{\Sigma-\Delta+a^2}
{\Sigma}a^2(\Sigma+a^2\sin^2 \theta)(2-\sin^2\theta)\big]\bigg\}K
 ,\\
\label{m1} && B^{\hat r} = \frac{B\csc\theta}{2K\Sigma}
\bigg[(\Sigma+a^2(2+\cos^2\theta)-\Delta)a^2\sin^2
\theta+\Sigma(\Sigma+a^2\sin^2
\theta)\nonumber\\
&&\qquad \quad-\frac{\Sigma-\Delta+a^2}
{\Sigma}a^2(\Sigma+a^2\sin^2 \theta)(2-\sin^2\theta)\bigg]
,\\
\label{m2} && B^{\hat\theta} =\frac{B\sin\theta\sqrt{\Delta}}
{K\Sigma} \bigg[r\Sigma+a^2(2-\sin^2\theta)
\frac{r\Delta-a^2r+(M-r) \Sigma} {\Sigma}\bigg] ,
\end{eqnarray}
which depend on angular momentum and tidal charge in complex way
and where we have used $K=
((\Sigma+a^2\sin^2\theta)^2-a^2\Delta\sin^2 \theta)^{1/2}$. In the
limit of flat spacetime, i.e. for $M/r\rightarrow 0$,
$Ma/r^2\rightarrow 0$ and $Q^*/r^2\rightarrow 0$, expressions
(\ref{e1})--(\ref{m2}) give the following limiting expressions: $
B^{\hat r}=B\cos\theta, B^{\hat\theta}=B\sin\theta, E^{\hat
r}=E^{\hat\theta}=0$, which coincide with the solutions for the
homogeneous magnetic field in Newtonian spacetime. {Here $\hat $
(hat) stands for orthonormal components of the electric and
magnetic fields.} Uniform magnetic field in the background of a
five dimensional black hole has been extensively studied in
~\cite{af04}. In particular authors presented exact expressions
for two forms of electromagnetic tensor and the electrostatic
potential difference between the event horizon of five dimensional
black hole and the infinity.

\section{\label{motion}
Charged Particle Motion in the Vicinity of Rotating Black Hole in Braneworld }

In this section we investigate in detail the motion of charged
particles around a rotating black hole in braneworld in an
external magnetic field given by 4-vector potential
(\ref{potentials}) with the aim to find a way for astrophysical
evidence for either the existence or nonexistence of tidal charge
$Q^*$. For simplicity of calculations we assume parameter $a$ to
be small, and obtain the exterior metric for slowly rotating
compact object in the braneworld in the following form
\begin{equation} \label{slowly_bran}
 ds^2=-A^2dt^2+H^2dr^2+r^2d\theta^2+r^2\sin^2\theta
d\varphi^2- 2\tilde{\omega}(r)r^2\sin^2\theta dt d\varphi\ ,
\end{equation}

here
\begin{equation}
A^2(r)\equiv\left(1-\frac{2M}{r}+\frac{Q^*}{r^2}\right)=H^{-2}(r),\
\ \ \ \ \ \ \ r>R\ ,
\end{equation}
is the Reissner-Nordstr\"{o}m-type exact solution~\cite{nkd00} for
the metric outside the gravitating object and $\tilde{\omega}(r) =
\omega(1-Q^{*}/2rM) =2Ma/r^3(1-Q^{*}/2rM)$.

The Hamilton-Jacobi equation
\begin{equation}
\label{Ham-Jac-eq} g^{\mu\nu}\left(\frac{\partial S}{\partial
x^\mu}+eA_\mu\right)\left(\frac{\partial S}{\partial
x^\nu}+eA_\nu\right)=-m^2\ ,
\end{equation}
for motion of the charged test particles with mass $m$ and charge
$e$ is applicable as a useful computational tool only when
separation of variables can be effected.

Since spacetime of the rotating object in braneworld  admits such
separation of variables (see e.g. \cite{dt02}) we shall study the
motion around source described with metric (\ref{slowly_bran})
using the Hamilton-Jacobi equation when the action $S$ can be
decomposed in the form
\begin{equation}
\label{action} S=-{\cal E}t+{\cal L}\varphi+S_{\rm
r\theta}(r,\theta)\ ,
\end{equation}
since the energy $\cal E$ and the angular momentum $\cal L$ of a
test particle are constants of motion in the spacetime
(\ref{slowly_bran}).

Therefore the Hamilton-Jacobi equation (\ref{Ham-Jac-eq}) with
action (\ref{action}) implies the equation for inseparable part of
the action as
\begin{eqnarray}
&& \frac{1}{2A^2}\left[{\cal E}+\frac{a}{r}  \left(\frac{2M\cal
L}{r^2}-\frac{Q^* {\cal L}}{r^3}+A^2eB\right)\right]\left[2{\cal
E}+aeB A^2 -aeB
\left(\frac{2M}{r}-\frac{Q^*}{r^2}\right)\sin^2\theta\right]+
\left({\cal L}+\frac{1}{2}eBr^2\sin^2\theta\right)\nonumber\\
&& \times \left[\frac{eB}{2}+\frac{\cal L}{r^2\sin^2
\theta}-\frac{a{\cal E}}{r^2
A^2}\left(\frac{2M}{r}-\frac{Q^*}{r^2}\right)\right]+A^2\left(\frac{\partial
S_{\rm r\theta}}{\partial
r}\right)^2+\frac{1}{r^2}\left(\frac{\partial S_{\rm
r\theta}}{\partial \theta}\right)^2=-m^2\ .
\end{eqnarray}

It is not possible to separate variables in this equation in
general case but it can be done for the motion in  the equatorial
plane $\theta = \pi/2$ when the equation for radial motion takes
form
\begin{equation}
\left(\frac{dr}{d\sigma}\right)^2={\cal E}^2-1-2V_{\rm eff}({\cal
E},{\cal L},r,\epsilon,a,Q^*)\ .
\end{equation}
Here $\sigma$ is the proper time along the trajectory of a
particle, ${\cal E}$ and ${\cal L}$ are energy and angular
momentum per unit mass $m$ and
\begin{eqnarray}
&& V_{\rm eff}({\cal E},{\cal L},r,\epsilon,a,Q^*)=
   \frac{a{\cal E}{\cal L}}{r^2}\left(\frac{2M}{r} -
   \frac{Q^*}{r^2}\right) + \left(\frac{{\cal
   L}^2}{2r^2}+\frac{\epsilon {\cal
   L}}{2}+\frac{\epsilon^2r^2}{8}+a{\cal E}\epsilon\right)\left
   (1-\frac{2M}{r}+ \frac{Q^*}{r^2}\right)
   -\frac{M}{r}+\frac{Q^*}{2r^2}\
\end{eqnarray}
is effective potential, where $\epsilon=eB/m$ is the magnetic
parameter.

\begin{figure*}
\includegraphics[width=0.45\textwidth]{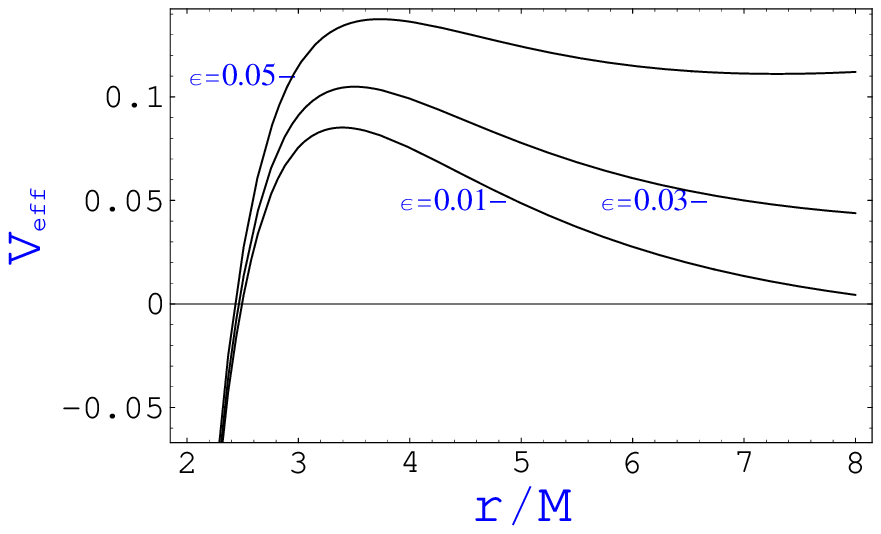}
\includegraphics[width=0.45\textwidth]{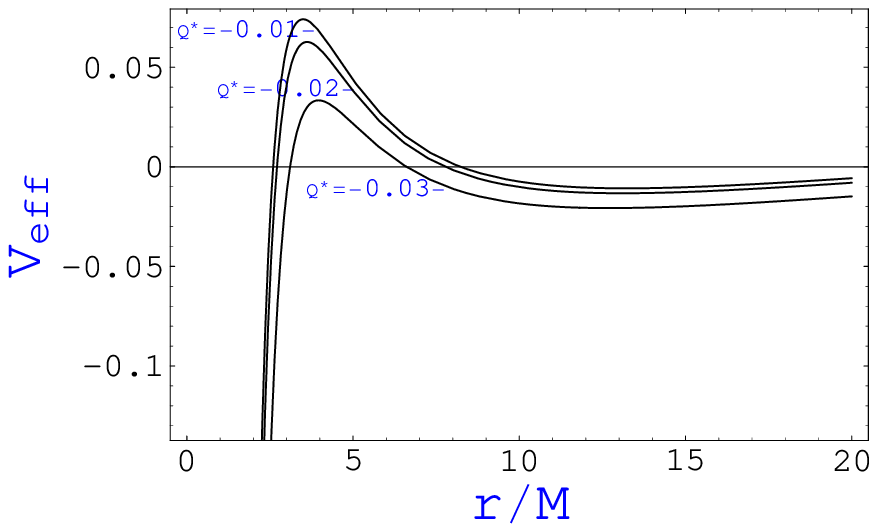}
\caption{\label{fig:1} Radial dependence of the effective
potential of the radial motion of the charged particles around
slowly rotating black hole in braneworld immersed in uniform
magnetic field for the different parameter of the magnetic field
$\epsilon$ (left graph) and tidal charge $Q^*$ (right graph) .}
\end{figure*}

Fig. \ref{fig:1} shows the radial dependence of effective
potential of the radial motion of charged particle on equatorial
plane of slowly rotating black hole in braneworld immersed in
uniform magnetic field for different values of  parameter of
magnetic field (left graph) and tidal charge (right one). One can
obtain now how magnetic and brane parameters change the character
of the motion of the charged particle. Both magnetic and tidal
parameters cause to shift the shape of the effective potential to
the observer in infinity that means the minimum distance of the
charged particles to the central object increases. As module of
the tidal charge increases parabolic and hyperbolic orbits start
to become unstable circular orbits, while magnetic parameter gives
opposite effect (Fig. \ref{fig:1}) (see e.g. our preceding
research~\cite{aak08}). Thus the radial profile of $V_{\rm eff}$
for different values of the tidal charge $Q^*$, running between
$-0.01$ and $-0.03$ shows that by increasing module of $Q^*$ from
$0.01$ to $0.03$ we also lower the potential barrier, as compared
to the Schwarzschild case, as expected for the potential of the
Reissner-Nordstr$\ddot{o}$m type black holes.

{The choice of the brane parameter's sign is stipulated according
to the following reason: the negative bulk cosmological constant
contributes to acceleration towards the brane, reflecting its
confining role on the gravitational field. In order for $U$  to
reinforce confinement, it must be negative. An effective energy
density $U=\kappa Q^*/r^4$ on the brane arising from the free
gravitational field in the bulk, where $\kappa$ is the positive
constant, needs not be positive. Indeed, $U < 0$ is the natural
case. In other words, negative tidal charge $ Q^*< 0$ is the
physically more natural case. Furthermore, $ Q^*< 0$ ensures that
the singularity is spacelike, as in the Schwarzschild solution,
whereas $Q^* > 0$ leads to a timelike singularity, which amounts
to a qualitative change in the nature of the general relativistic
Schwarzschild solution (see for more details~\cite{nkd00}).}

\section{\label{branemotion}Motion of Test Particle Around Black
Hole in Braneworld}

In order to find exact analytical solution for radius $r_{\rm
ISCO}$ we assume that external magnetic field is absent and black
hole in braneworld is nonrotating when metric (\ref{slowly_bran})
can be written in the diagonal form as
\begin{equation}
ds^2=-\frac{\Delta}{r^2}dt^2+\frac{r^2}{\Delta}dr^2+ r^2 d\theta^2
+ r^2\sin^2\theta d\varphi^2\ ,
\end{equation}
where $\Delta=r^2-2Mr+Q^{*}$ does not include terms being
proportional to angular momentum of black hole. Now using the
Hamilton-Jacobi method described in previous section \ref{motion}
one can easily find equation of motion of test particle in the
equatorial plane of the black hole in braneworld as
\begin{eqnarray}
&& \label{tuch}\frac{dt}{d\sigma}={\cal E}\frac{r^2}{\Delta}\ ,\\
&& \label{ruch}\left(\frac{dr}{d\sigma}\right)^2={\cal
E}^2-\frac{\Delta}{r^2}
\left( 1+\frac{{\cal L}^2}{r^2} \right)\ ,\\
&&\label{fuch}\frac{d\varphi}{d\sigma}=\frac{{\cal L}}{r^2}\ .
\end{eqnarray}
Using the equations (\ref{ruch}) and (\ref{fuch}) and introducing
new variable $u=1/r$ one can obtain the following equation
\begin{equation} \label{polyn}
\left(\frac{du}{d\varphi}\right)^2=-Q^{*}u^4+2Mu^3-\left(1+\frac{Q^{*}}{{\cal
L}^2 }\right)u^2+\frac{2M}{{\cal L}^2}u-\frac{1-{\cal E}^2}{{\cal
L}^2}=f(u)\ ,
\end{equation}
which defines the trajectory of the test particle around black
hole in braneworld. The condition of occurrence of circular orbits
is:
$$
f(u)=0\,,\ f'(u)=0\ .
$$
From these equations, it follows that energy ${\cal E}$ and
angular momentum ${\cal L}$ of a circular orbit of radius
$r_{c}=u_{c}$ is given by
\begin{eqnarray}
&& \label{encirc}{\cal E}^2=\frac{(1-2Mu+Q^{*}u^2)^2}{1-3Mu+2Q^{*}u^2}\ ,\\
&& \label{enmomn}{\cal L}^2=\frac{M-Q^{*}u}{2Q^{*}u^3-3Mu^2+u}\ .
\end{eqnarray}
\begin{figure*}
\includegraphics[width=0.45\textwidth]{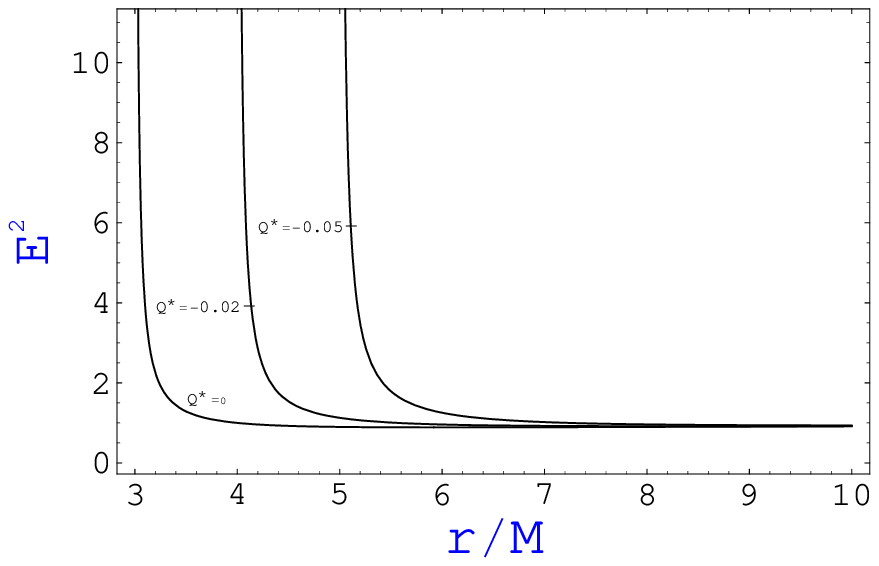}
\includegraphics[width=0.45\textwidth]{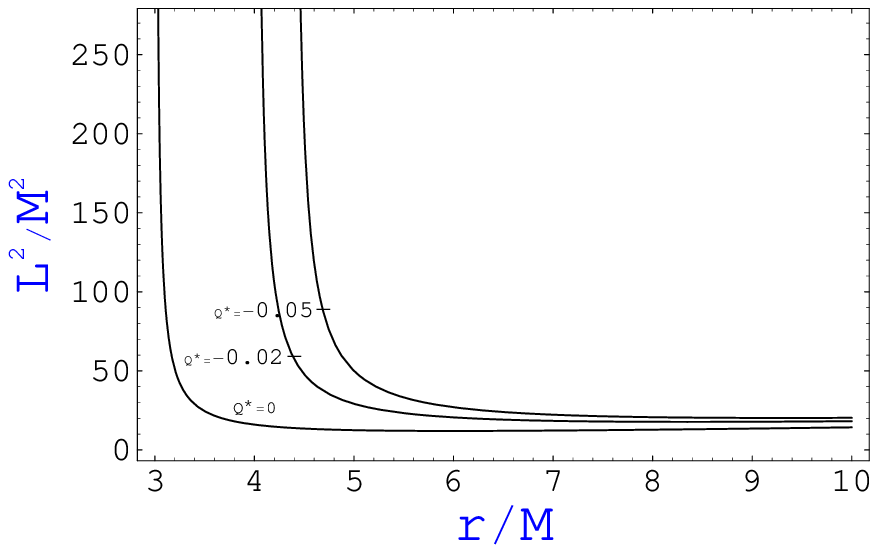}
\caption{\label{fig:enandmom} {Radial dependence of energy (left
graph) and angular momentum (right graph) of circular orbits
around black hole in braneworld for the different values of the
brane tension $Q^*$. For comparison we have also plotted the Schwarzschild dependence,
corresponding to $Q^* = 0$.}}
\end{figure*}
{Fig. \ref{fig:enandmom} shows the radial dependence of both the
energy and the angular momenta of the test particle moving on
circular orbits in the equatorial plane. One can easily see that
presence of the brane parameter forces test particle to have
bigger energy and angular momentum in order to be kept on its
circular orbit. It is a consequence of the increase of the
gravitational potential of the central object in braneworld.}

From equations (\ref{encirc}) and (\ref{enmomn}) one can easily
find minimum radius for circular orbits $r_{\rm mc}$
\begin{equation}
r_{\rm mc}>\frac{4Q^{*}}{3M-\sqrt{9M^2-8Q^{*}}}\ ,
\end{equation}
or if we expand this expression in degrees of $Q^*/M^2$,
it takes the following form:
\begin{equation}
  r_{\rm mc}\thickapprox 3M-\frac{2Q^*}{3M} -\frac{4 Q^{*2}}{27M^3}
+{\mathcal{O}}\left(
  \frac{Q^{*3}}{M^5}\right)\ .
\end{equation}

In the limiting case when $Q^{*}$ tends to zero $r_{\rm mc}=3M$
which coincides with the Schwarzschild limit.
The minimum radius for a stable circular orbit will occur at point
of inflexion of the function $f(u)$, or in other words we must
supplement conditions $f(u)=f'(u)=0$ with the equation $f''(u)=0$.
Then one can easily obtain the equation
\begin{equation}
4Q^{*2}u^3-9MQ^{*}u^2+6M^2u-M=0\ ,
\end{equation}
and its solution in the form
\begin{equation}
\label{isco}
r=\frac{4Q^{*}}{3M+\sqrt[3]{A-B}+\sqrt[3]{A+B}}\equiv
r_{\rm ISCO}\ ,
\end{equation}
where
\begin{equation}
A=8MQ^{*}-9M^3\,,\ \, B=4\sqrt{(4MQ^{*}-5M^3) (MQ^{*}-M^3)}\ ,
\end{equation}
or if we expand this expression in degrees of $Q^*/M^2$,
it takes the following form:
\begin{equation}
  r_{\rm ISCO}\thickapprox 6M-1.5\frac{Q^*}{M} +0.0078
  \frac{Q^{*2}}{M^3}+{\mathcal{O}}
  \left(\frac{Q^{*3}}{M^5}\right)\ .
\end{equation}

To the best of our knowledge the analytical
expression~(\ref{isco}) is original one. It defines the limit of
the stability of innermost circular orbit in vicinity of black
hole in braneworld. Numerical solutions with similar results for
$r_{\rm ISCO}$ around rotating black hole in braneworld and
circular orbits in accretion disks have been studied in papers
\cite{ag05} and \cite{harko08}, respectively.

\begin{figure}
\includegraphics[width=0.45\textwidth]{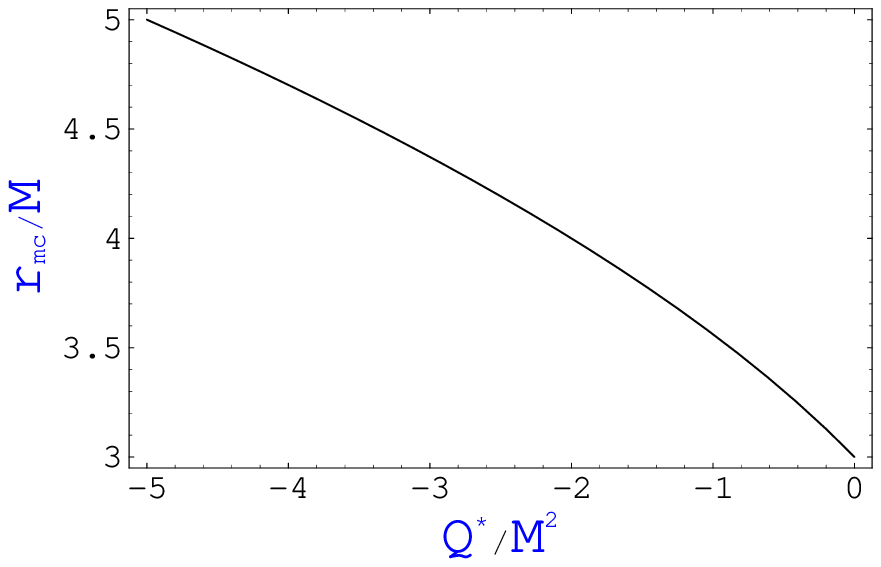}
\includegraphics[width=0.45\textwidth]{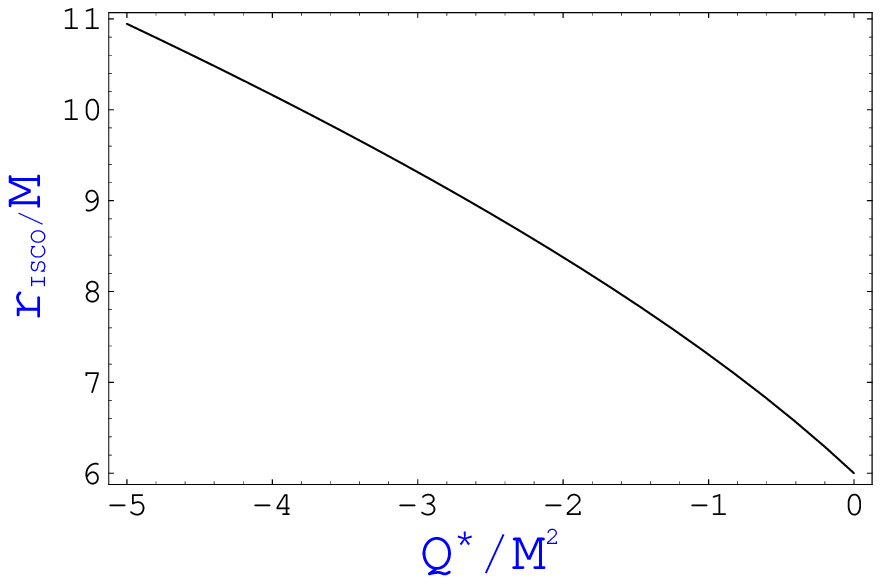}
\caption{\label{circisco} {Dependence of the lower limit for
radiuses of circular orbits $r_{m c}$ (left graph) and ISCO
$r_{\rm ISCO}$ (right graph) from the tidal charge $Q^*$.}}
\end{figure}
{The dependence of the minimum radius for circular orbits $r_{\rm
mc}$ and radius of ISCO around black hole from the brane tidal
charge is plotted in the Fig. \ref{circisco}, where the values
related to the Schwarzschild black hole correspond to $Q^* = 0$.
One can easily see from the plots that presence of the tidal
charge forces the radius of the stable orbits to be shifted away
from the central object in the direction of an observer at
infinity which confirms the earlier results of Aliev \&
G\"{u}mr\"{u}k\c{c}\"{u}o\v{g}lu \cite{ag05}.}

The variation of $Q^*$ also modifies the position of the
marginally stable orbit, as shown by the shift of the ISCO, which
is presented in the left plot in the Fig. \ref{circisco}. The
negative decreasing charges lead to the increase of ISCO radius.
By decreasing the value of $Q^*$ from 0 to -5, we shift the radius
of ISCO  to bigger and bigger values. The lower values of the
potential for $Q^*$ involve a lower specific energy of the
orbiting particles. As we decrease $Q^*$ from 0 to -5, ISCO radius
is increasing from values greater than the radius of the
marginally stable orbit for the Schwarzschild geometry to bigger
ones. The efficiency has an opposite trend with compare to angular
momentum: for negative tidal charges it has bigger values than in
the case of the Schwarzschild black holes.

Next, we will give clear derivation of the capture cross section
of slowly moving test particles by black hole in braneworld (Slow
motion means that ${\cal E}\simeq 1$ at the infinity.).  The
critical value of the particle's angular momentum, ${\cal L}_{\rm
cr}$, hinges upon the existence of a multipole root of the
polynomial $f(u)$ in (\ref{polyn}) \cite{zakhar}. For convenience
hereafter we rewrite the equation (\ref{polyn}) in terms of
dimensionless parameters as radial coordinate $r\rightarrow r/M$,
momentum  ${\cal L}\rightarrow {\cal L}/M$ and tidal charge
$Q^*\rightarrow Q^*/M^2$:
\begin{equation}
\label{pleq}
r^3-\frac{{\cal L}^2+Q^*}{2}r^2+{\cal L}^2r-\frac{Q^*{\cal
L}^2}{2}=0\ .
\end{equation}
Cubic equation (\ref{pleq}) has a multiple root if and only if its
discriminant vanishes. After simple algebraic transformations one
can easily obtain the following equation for particle angular
momentum
\begin{equation} \label{aneq}
{\cal L}^6 (1-Q^*)-{\cal L}^4(3Q^{*2}-20Q^*+16)- {\cal L}^2 Q^{*2}
(8+3Q^*)-Q^{*4}=0\ ,
\end{equation}
{which has an exact solution in the form}
\begin{equation}  \label{momencr}
{\cal L}_{{\rm cr}}^2= \left\{
 \begin{array}{lcl} 
    \sqrt[3]{-B_{\rm 1}/2+\sqrt{D}}+\sqrt[3]{-B_{\rm 1}/2-\sqrt{D}}-\frac{(20Q^*-3Q^{*2}-16)^2}
    {3 (1-Q^*)}\ ,\\
    2 \sqrt{\frac{-A_{\rm 1}}{3}}\cos\left\{\frac{1}{3}\arccos\left[-B_{\rm 1}/(2
    \sqrt{-(A_{\rm 1}/3)^2})\right]\right\}-\frac{(20Q^*-3Q^{*2}-16)^2}
    {3 (1-Q^*)}\ ,\\
 \end{array}
 \begin{array}{rcl}  
    D\geq 0 \ ;\\
    \\
    D<0\ .\\
 \end{array}
\right.
\end{equation}
{Here we have introduced the following notations}
\begin{eqnarray}
&& A_{\rm 1}=-\frac{(20Q^*-3Q^{*2}-16)^2}{3 (1-Q^*)^2}-\frac{8
Q^{*2}+3Q^{*3}}{1-Q^*}\ ,\nonumber\\
&& B_{\rm 1}=2\frac{(20Q^*-3Q^{*2}-16)^2}{27 (1-Q^*)^3}-\frac{(20
Q^*-3Q^{*2}-16)^2(8 Q^{*2}+3 Q^{*3})}{1-Q^*}-\frac{Q^{*4}}{1-Q^*}\
,\nonumber \\
&& D=\frac{A_{\rm 1}^3}{27}+\frac{B_{1}^2}{4} \ .\nonumber
\end{eqnarray}
{In the limiting case, i.e. when tidal charge vanishes the
solution of the equation (\ref{aneq}) is ${\cal L}=4$, which
coincides with critical angular momentum for particle capture
cross section for Schwarzschild black hole \cite{mtw}. As a particle having a
critical angular momentum travels from infinity towards the black
hole in braneworld, it spirals into an unstable circular orbit of
radius given as}
\begin{equation}
r_{\rm uc}=2\sqrt[3]{\left( \frac{{\cal L}^2+Q^*}{6}\right)^3- {\cal
L} \left( \frac{{\cal L}^2+Q^*}{6}\right)+{\cal L}^2 Q^*} +
\frac{{\cal L}^2+Q^*}{6}\ .
\end{equation}

\begin{figure}
a) \includegraphics[width=0.45\textwidth]{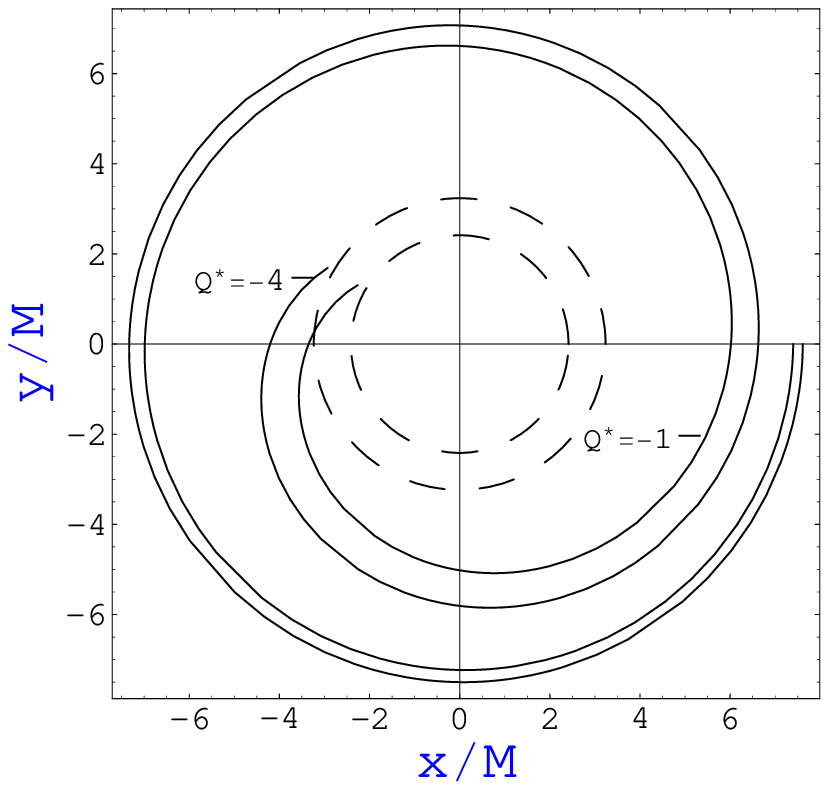}
b) \includegraphics[width=0.45\textwidth]{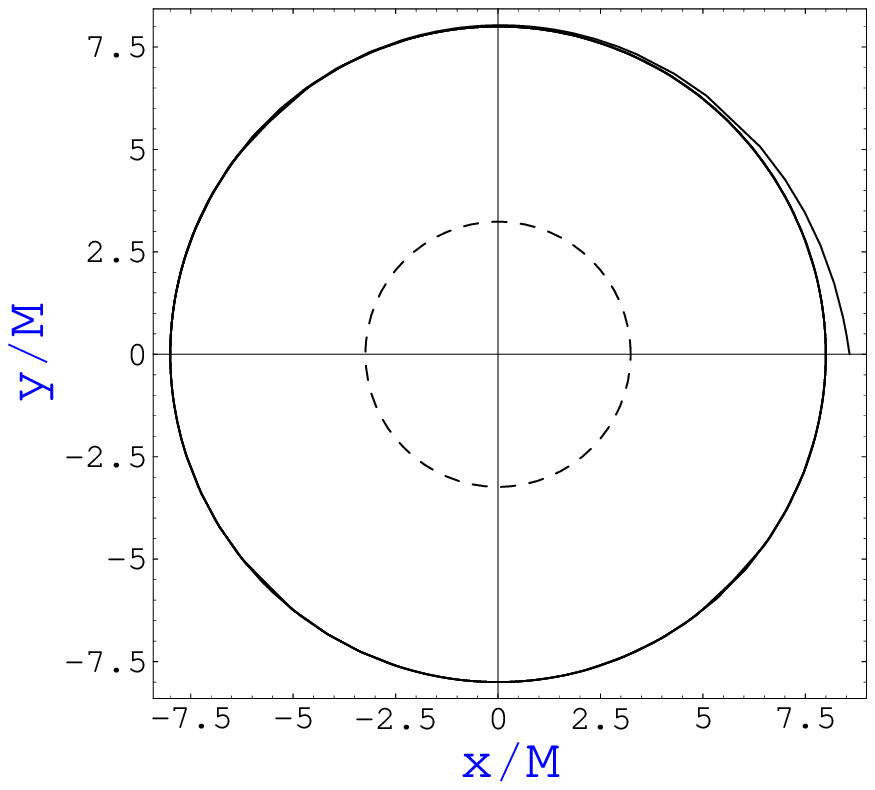}
\caption{\label{trajectories} a). Orbits of the test particle
falling into central black hole for different values of the tidal
charge $Q^*$. b). Stable circular
orbit of the test particle around black hole in braneworld. In all
plots horizons are shown with dash lines.}
\end{figure}

Finally in Fig. \ref{trajectories} we present the shapes of
different kinds of trajectories of test particles around black
hole in braneworld, which are given by equation (\ref{polyn}). The
trajectories of test particles falling to the central black hole
in braneworld for different values of the brane parameter are
shown in Fig. \ref{trajectories} a). From the plot one can obtain
that increase of the module of the brane parameter causes orbits
shift to an observer at the infinity, which is a consequence of
increase of the radius of the event horizon by braneworld effects.
Fig. \ref{trajectories} b) illustrates the sample of unstable
circular orbits of the particles, while Fig. \ref{trajectories} c)
shows the shape of the circular orbits around black hole in the
braneworld.

\section{\label{conclusion} Conclusion}

We have concentrated here on the basic physical properties of
particle motion and magnetic field in the background spacetime
metric of the braneworld black holes. Motivation of this research
is caused by the fact that testing strong field gravity and the
detection of the possible deviations from standard general
relativity, signaling the presence of new physics, remains one of
the most important objectives of observational astrophysics.
Because of their compact nature, black holes provide an ideal
environment to perform precise relativistic measurements, in
particular the observational possibilities for testing the DMPR
solution of the vacuum field equations in brane world models.

Here the physical parameters of the effective potential, and ISCO
have been explicitly obtained for several values of the parameters
characterizing the vacuum DMPR solution of the field equations in
the braneworld models. We have found original exact expression for
the lower limit of innermost stable circular orbits of test
particle around black hole in braneworld (Before ISCO behavior in
braneworld models has been investigated only
numerically~\cite{ag05,harko08}.). Then we have plotted the
dependence of the ISCO radius from the brane tidal charge and
particle trajectories around black hole in braneworld.

The best constraints on the braneworld black hole parameters were
recently obtained from the classical tests of general relativity
(perihelion precession, deflection of light, and the radar echo
delay, respectively)~\cite{bhl08}. The existing observational
solar system data on the perihelion shift of Mercury, on the light
bending around the Sun (obtained using long-baseline radio
interferometry), and ranging to Mars using the Viking lander, were
applied to the relativistic in  DMPR spacetime, can constrain the
numerical values of the brane parameter. The strongest limit
$|Q^*|\lesssim 10^8 {\rm cm}^2$ was obtained from the Mercury's
perihelion precession.

The recent measurements of the ISCO radius in accretion disks
around black holes may also give alternate constraints on the
numerical values of the brane tidal charge.  All the astrophysical
quantities related to the observable properties of the accretion
disk can be obtained from the black hole metric and observations
in the near infrared  or X-ray bands have provided important
information about the spin of the black
holes~\cite{narayan06,narayan08}. It was stated that rotating
black holes have spins in the range $0.5 \lesssim a \lesssim 1$
that is according to the observations ISCO radii are essentially
shifted towards the central objects and there is no any effect
measured from the brane tidal charge which acts in the opposite
direction.

Because of the differences in the spacetime structure, the brane
world black holes present some important differences with respect
to their disc accretion properties, as compared to the standard
general relativistic Schwarzschild and Kerr cases. Therefore, the
study of the innermost stable orbits in the vicinity of compact
objects is a powerful indicator of their physical nature. Since
the ISCO radius in the case of the braneworld black holes is
different as compared to the standard general relativistic case,
the astrophysical determination of these physical quantities could
discriminate, at least in principle, between the different gravity
theories, and give some constrains on the existence of the extra
dimensions. Finally, since there was no braneworld effect on
stable orbits around black holes on the scale of $a$ order of
$10^8 {\rm cm}^2$, we may conclude that from astrophysical point
of view on the base of comparison of observations of ISCO in
accretion disks around black holes and ISCO analysis around black
hole in braneworld that brane tidal charge has an upper limit
$\lesssim 10^9 {\rm cm}^2$. We roughly estimated that one order
less magnitude of $Q^*$ may not affect on the observational data
on ISCO data around black holes.

\begin{acknowledgments}

{Authors gratefully thank Naresh Dadhich for his invaluable help,
extensive discussions, editing the text and making important
corrections and comments.} They also thank Valeria Kagramanova for
useful discussions. Authors thank the IUCAA for warm hospitality
during their stay in Pune and AS-ICTP for the travel support
through BIPTUN program. This research is supported in part by the
UzFFR (projects 5-08 and 29-08) and projects FA-F2-F079 and
FA-F2-F061 of the UzAS. This work is partially supported by the
ICTP through the OEA-PRJ-29 project and the Regular Associateship
grant.

\end{acknowledgments}

\end{document}